\documentclass[preprint,onecolumn,amsmath,superscriptaddress,amssymb,floatfix,showkeys]{revtex4}
\usepackage[dvips]{graphicx}
\usepackage{dcolumn}

\usepackage{xspace}
\def\etc{etc.\@\xspace}
\def\ie{i.e.\@\xspace}

\begin{document}
\title{Pore-size dependence and characteristics of water diffusion in slit-like micropores}
\author{S.O. Diallo}
\affiliation{Chemical and Engineering Materials Science Division,  Oak Ridge National Laboratory, Oak Ridge, Tennessee 37831, USA}



\begin{abstract}
The temperature dependence of the dynamics of water inside microporous activated carbon fibers (ACF) is investigated by means of incoherent elastic and quasi-elastic neutron scattering techniques. The aim is to evaluate the effect of increasing pore size on the water dynamics in these primarily hydrophobic slit-shaped channels. Using two different micropore sizes ($\sim$12 and 18 {\AA}, denoted respectively ACF-10 and ACF-20),  a clear suppression of the mobility of the water molecules is observed as the pore gap or temperature decreases. This suppression is accompanied by a systematic dependence of the average translational diffusion coefficient $D_r$ and relaxation time $\langle\tau_0\rangle$ of the restricted water on pore size and temperature. The observed $D_r$ values are tested against a proposed scaling law, in which the translational diffusion coefficient $D_r$ of water within a porous matrix was found to depend solely on two single parameters, a temperature independent translational diffusion coefficient $D_c$ associated with the water bound to the pore walls  and the ratio $\theta$ of this strictly confined water to the total water inside the pore, yielding unique characteristic parameters for water transport in these carbon channels across the investigated temperature range.
\end{abstract}

\maketitle

\date{\today}

Due to its polar nature, water is considered to be the {\it de-facto}  solvent of choice for many important chemical and biological processes. In the animal kingdom for example, water facilitates key essential functions such as the regulation of body temperature, the breaking down of nutrients, the activities of vital protein and enzymes, to name a few.   These vital roles coupled to the extraordinary thermo-physical properties of water, such as its ability to expand rather than contract upon cooling,  and its known thermodynamics anomalies (heat capacity, and isothermal compressibility \etc) continue to drive fundamental research on this complex fluid.

In many common life circumstances however, water is often constrained to some surfaces, or restricted within tight cavities or very small voids or cracks. This nano-scopically restricted water undergoes structural perturbations and exhibits reduced molecular mobility and thermodynamics properties that are very distinct from those of bulk water \cite{Teixeira:85,Stanley:99, Gelb:99,Alba-Simionesco:06, Chaplin:09,Alabarse:12}.   The slowing down of water dynamics in nano-confinement is rather well documented, and has been observed with neutron spectroscopy in a number of  porous materials, including the hydrophilic silica-based systems such as GelSil \cite{Crupi:03}, Vycor \cite{Zanotti:99a}, SBA-15 \cite{Webber:04}, MCM-14 \cite{Liu:04,Mansour:02,Liu:06,Takahara:05}, and FSM-12 \cite{Diallo:12}, and the hydrophobic carbon systems such as carbide-derived carbon (CDC) \cite{Chathoh:11}, single-wall nanotubes (SWNT) \cite{Mamontov:06}, and double-wall nanotubes (DWNT) \cite{Mamontov:06,Chu:07}.  

The most commonly investigated pore geometry with neutrons in confined water research is that of the cylindrical type (\ie silica MCM-41 or carbon SWNT), which is comparatively easier to model than non-uniform or other complex shapes. This convenient geometry being readily available in various sizes, it facilitates systematic studies and enables direct comparison with proxy models used in molecular dynamics simulations \cite{Liu:04,Bourg:12}. 

In reality though, natural confinement of water occurs under various spatial and geometrical restrictions.  Understanding the effects of other pore shapes and spacings on the dynamics of water is thus of key scientific importance. In the present study, we investigate the effect of pore size on the dynamics of water confined inside carbon channels that are primarily slit-shaped.  Specifically,  the characteristic diffusive dynamics of water confined in the micropores of two activated carbon fiber samples have been investigated  at various temperatures using neutron spectroscopy.   The observed average characteristic relaxations $\langle \tau_0\rangle$  and translational diffusion coefficients $D_r$  (associated with the  H-sites of water) indicate a diffusion that is remarkably slower than in the bulk liquid, with an Arrhenius temperature behavior and characteristic energy  barrier $E_A$ that is however similar to the bulk value.  The diffusion properties can be adequately parametrized using a recently proposed scaling law \cite{Chiavazzo:14} for water transport, leading to two unique water transport  characteristics in these slit-like carbon pore structures. These observed parameters can be used to predict the translational diffusion coefficient of water in these porous carbon samples at any other temperature where the law holds, without having to perform additional measurements.  

\section{Sample Characteristics and Details}\label{sec:sample}

The Kynol\texttrademark $^*$ activated carbon fiber (ACF) samples were received from the American Technical Trading (ATT), NY.  These ACF samples were synthesized from polymeric carbon precursors and contain narrow pore size distributions and a large pore volume that is set by the degree of activation during synthesis.  We use two different ACF samples  (ACF-1603-10, and ACF-1603-20) for our study.  The as-received carbon fibers have average macroscopic dimensions of  $\sim$ 3 mm length, and $\sim$ 10 $\mu$m diameter. These nominal average dimensions were checked against transmission electron microscopy results for consistency (not shown here).  Previously reported scanning tunneling microscopy (STM) of similar ACF samples  \cite{Kaneko:92a,Daley:96,Parades:01} has revealed a highly sinuous pore network with predominantly uniform nanometer size pore distribution (with a micropore volume $V_{\mu pore}$ of $\sim$88\% of the total pore volume [$V_{Total}\sim$0.4-0.5 cc/g] in ACF-10  \cite{Daley:96,LiuW:14}). The remainder of the pore volume consists of random meso-pores and some ultra-micropores. The relevant sample characteristics are listed in Table \ref{tbl1}. Various other measurements including thermodynamics\cite{Kaneko:92a, Kaneko:92b}, small-angle neutron and X-ray scattering \cite{Suzuki:92,Ramsay:98} of ACF-10 and ACF-20 have also confirmed the interconnected pore structure, made primarily of elongated curvy slit-pores.  The pores themselves are the voids between curvy but parallel carbon sheets \cite{Hayes:14}.  The gap between two non-flat  carbon sheets defines the average pore size.  The key point is that the pores in ACF are not of cylindrically shaped as in the other well characterized nanotubes, but appear to made of irregular slits pores, as indicated by the above cited STM work. This different pore structure offers a new platform for investigating fluids in confined geometries in non-cylindrical pores. For the current ACF-10 and ACF-20 samples, the average pore size is estimated to be respectively (12$\pm0.5$) and (18$\pm$0.6) {\AA}, based of previously reported N$_2$ adsorption isotherm measurements  \cite{Daley:96,LiuW:14} using the Dubnin-Astakhov equation \cite{Dubinin:89}. Since the average pore sizes are all below 2 nm, we refer to the samples as microporous materials.  The nominal sample specific surface areas, tabulated in Table \ref{tbl1},  were  confirmed at the time of the synthesis by the manufacturer by iodine number testing, and correlated well with BET calculations in this range \cite{Daley:96}. Iodine is a common standard adsorbate used in industry to estimate the adsorption capacity of carbon samples. 

\begin{table}
\caption{Characteristic size and volume of the micropores in ACF, corresponding BET surface area (from Ref.  \cite{LiuW:14}) and hydration level $h$ (weight\%), as measured from the relative weight change of \lq dry'  sample after 24 hours exposure to a  humid atmosphere. $V_{\mu pore}$ is the open volume of micropores, and $V_{Total}$ the total available pore volume in the sample.  \label{tbl1}}
\begin{ruledtabular}
\begin{tabular}{c | c | c | c | c}
Sample	& $\mu$-pore size ({\AA})	&  $\frac{V_{\mu pore}}{V_{Total}}$ (\%) & Surface Area (m$^2$/g) & $h$ (weight\%)\\
\hline
ACF-10	&11.97	& 88	&  785	& 21 \\
\hline
ACF-20	&17.69	& 75	 & 2247	 & 23 
\end{tabular}
\end{ruledtabular}
\end{table}

To prepare for the neutron experiments, we outgassed the as-provided ACF samples for 36 hours at 473 K in a vacuum oven to remove all of the bulk-like water, and most of the surface water that was originally present in the as-received sample. We  then exposed respectively about 2 g of each so-dried sample to a humid atmosphere in a desiccator for several hours. The hydration level reached in each case after about 24 hours exposure  is indicated in Table \ref{tbl1}. These values were based on the relative weight change of each sample. These somewhat important hydration levels signal the presence of  hydrophilic groups (such as oxygenated sites for example) in otherwise totally hydrophobic samples. Since the quoted hydration amounts  are relative to the \lq drying' conditions set above, it is important to note that the diffusive dynamics reported in this work are those of all water molecules present inside the porous carbon network.  The neutron being primarily sensitive to hydrogen atoms, the present  measurements yield the characteristic relaxations of all confined water molecules.  The hydrated samples were subsequently each loaded unto two concentric Al cylinders (with a 2 mm gap between them) to minimize multiple scattering. The subsequently indium-sealed containers were anchored to the copper finger of a close-cycle refrigerator (CCR) stick which allowed to control the sample temperature for the present measurements between 50 and 300 K.

\begin{figure*}
\includegraphics[width=0.9\linewidth]{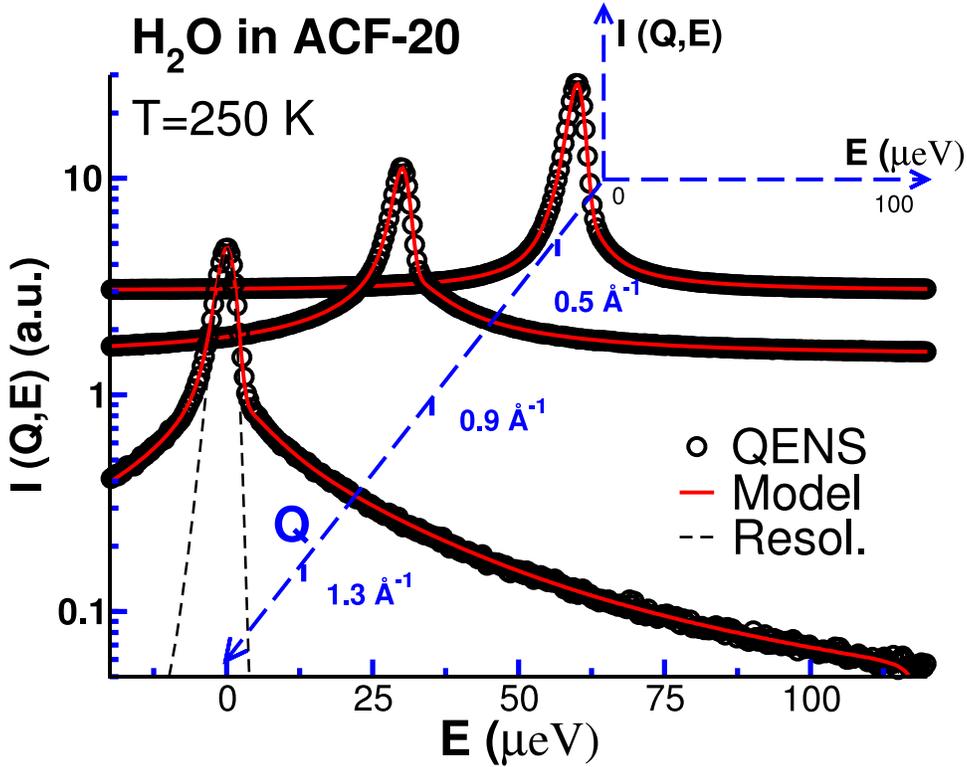}
\caption{(Color online) Observed quasi-elastic neutron scattering intensity $I(Q,E)$ on the BASIS instrument, as a function of energy transfer $E$ for water in ACF-20 (average slit size $d\simeq18${\AA}) at selected momentum transfer $Q$ and temperature $T=250$ K. Although the actual energy  window investigated is symmetric in energy ($\pm$120 $\mu$eV), only the neutron energy gain side is shown for display purposes. The relative errorbars are smaller than the symbols.}
\label{fig.1}
\end{figure*}

\begin{figure}
\includegraphics[width=0.9\linewidth]{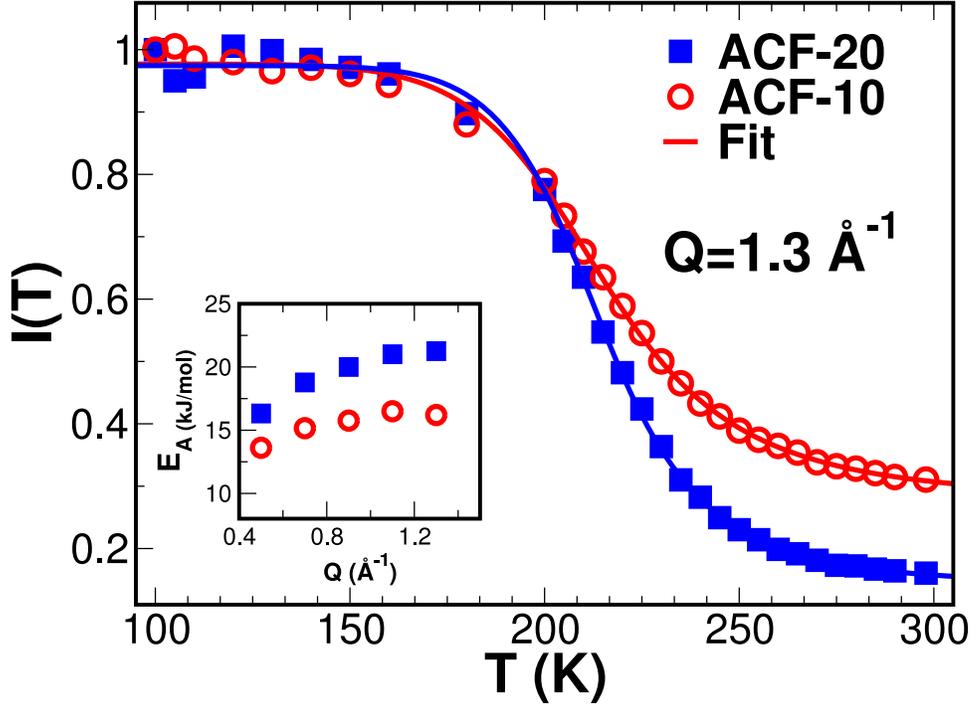}
\caption{(Color online) Temperature and pore size dependence of the energy-integrated elastic intensity $I(T)=\int_{-\Gamma_r}^{\Gamma_r} dE ~ S_T(Q,E)$ of water confined in carbon fibers, at a selected momentum transfer $Q$ of 1.3 {\AA}, where $\Gamma_r$ is the instrument energy resolution width (HWHM=$1.75\mu$eV). The onset of observable diffusive dynamics on BASIS is marked by the departure from a monotonically varying slope in $I(T)$ at low temperatures to a more rapidly changing slope at the higher temperatures. The  solid lines are model fits to data, corresponding to an activation energy of $E_A=$16 and 21 kJ/mol, for ACF-10 and ACF-20 respectively. The errorbars are smaller than the symbols.}
\label{fig.2}
\end{figure}

 \section{Neutron Scattering Measurements}\label{sec:meas}  
In this section, the technical details concerning the neutron measurements are presented.  The neutron data were all collected on the backscattering spectrometer (BASIS) at Oak Ridge National Laboratory (ORNL), USA \cite{Mamontov:11}. This unique indirect geometry neutron spectrometer has an excellent energy resolution $\Gamma_r=$ 1.75 $\mu$eV (Half-Width-at-Half-Maximum or HWHM) at the elastic line, and covers a momentum $Q$ and energy $E$ transfer range, respectively 0.3 $\le Q\le$ 2 {\AA}$^{-1}$, and  $-$120 $\le E\le$ 120  $\mu$eV, enabling access to time and length scales in the range $l=$3-22 {\AA} and $t$ from $\sim$5 up to 1000 ps.  To show the quality of the data collected on BASIS, representative $S(Q,E)$ spectra  of water adsorbed in ACF-20 at temperature 250 K is shown in Fig. \ref{fig.1}. 

\subsection{Incoherent Elastic Response}
To gain preliminary insights into the molecular dynamics of water in ACF and determine the appropriate temperature range within which the diffusive dynamics become observable on BASIS,  it is useful to investigate the temperature dependence of the elastic peak. This is achieved by integrating the peak intensity of rapidly collected  runs (10 mins per point) of $S(Q,E)$ around $E=0$ ($-\Gamma_r\le E\le \Gamma_r$) at each temperature from say 100 to 300 K, in steps of 5-10 K depending on the temperature region. Fig. \ref{fig.2} shows the result of such an analysis for both ACF samples, at a selected $Q$ of 1.3 {\AA}. The resulting energy-integrated elastic intensity $I(T)$ is normalized to the lowest temperature data taken at $T_0=50$ K so that $I(T)/I(T_0)$ equals unity at $T=T_0$. The $I(T)$ signal decreases with increasing temperature, indicating an increase in mobility of the H-sites of water, and consequently of water itself.   This behavior can be easily understood on the basis of an $I(T)$ that is modulated by a Debye-Waller coefficient; \ie proportional to $\exp{[-Q^2\langle u^2(T)\rangle/3]}$ \cite{Chen:99,Magazu:08} at each $Q$, where $\langle u^2(T)\rangle$ is the mean square displacement (MSD) associated with the hydrogen atoms in water. At low $T$, the MSD arises primarily from harmonic vibrations, increasing monotonically with increasing $T$.  At high $T$ where anharmonic vibrations become important, its accurate determination requires a subtraction of the harmonic phonon contributions  (which can be estimated from the MSD at low $T$ by extrapolation). 

In any case, the total elastic area $I(T)$ integrated over  the instrument resolution window  would effectively increase as $T$ is reduced, converging to some constant value  when the diffusive dynamics become resolution limited, which is observed here to be around 150-180 K. Similarly, when the dynamics become too broad to be  observed at the high temperatures (above $\sim$ 290 K here),  $I(T)$ flatten to background levels. These levels are dominated by higher energy dynamical processes that fall outside the instrument dynamics range. The intermediate region between these two limits where the intensity drops relatively quickly is associated with the relevant molecular diffusion. Water crystallization on the other hand, if present, would manifest itself along the $I(T)$ curve as a sharp drop (similar to a step-like first order transition), which is not observed here at any $Q$ (\ie at any length scale probed). The scattering from solid ice is purely elastic. Our present observation suggests that water molecules in the microporous ACF remain mobile well below bulk water freezing temperature.

To get a preliminary estimate of the energies associated with the thermally activated water dynamics suggested by the drop in  $I(T)$ at various length scales,  the diffusive component of $S(Q,E)$ can be modeled with a single Lorentzian function $L_T(E)=\frac{1}{\pi}\frac{\Gamma_T}{E^2+\Gamma_T^2}$ to a first approximation. Assuming an Arrhenius temperature dependent  $\Gamma_T=\Gamma_\infty\exp(-E_A/RT)$), the energy-integrated $I(T)$  for $E\le \Gamma_r$  can be explicitly evaluated at each $Q$ as follows,  \cite{Springer:77, Grapengeter:87},
\begin{eqnarray}
I(T)&=&\int_{-\Gamma_r}^{\Gamma_r}dE ~S(Q,E) \nonumber \\
&=&\int_{-\Gamma_r}^{\Gamma_r}{dE \left[A_0\delta(E)+(1-A_0)\frac{1}{\pi}\frac{\Gamma_T}{E^2+\Gamma_T^2}  \right]} \nonumber\\
&= &  A_0+(1-A_0)\frac{2}{\pi}\arctan\left(\frac{\Gamma_r}{\Gamma_T}\right) \nonumber\\
&= &A_0+(1-A_0)\frac{2}{\pi}\arctan\left(\frac{\Gamma_r}{\Gamma_\infty e^{-E_A/RT}}\right)
\label{eq:EIS}
\end{eqnarray}
\noindent where $\Gamma_r$ is the HWHM of the instrument energy resolution ($\sim$1.75 $\mu$eV), $\Gamma_\infty$ that of $L_T(E)$ when $T\rightarrow\infty$, and $A_0$ a fraction of the observable molecules that do not participate to the diffusive process.   Eq. \ref{eq:EIS} has thus three adjustable parameters: $A_0$, $\Gamma_\infty$, and $E_A$. The solid lines in Fig. \ref{fig.4} represent the fits obtained at $Q=$1.3 {\AA}$^{-1}$ using Eq. \ref{eq:EIS}. These fits yield average $E_A$ values between 13 and 16 kJ/mol for ACF-10 and 16 and 21 kJ/mol for ACF-20 for the $Q$-range probed (0.5 $\le Q\le 1.3$ \AA$^{-1}$).   At the higher $Q$ ($\ge1.5$ {\AA}$^{-1}$),  the data analysis is complicated by the coupling between translational and rotational modes. These $Q$ values are thus not included in the present analysis.  Estimate of the energy barriers $E_A$  associated with the long range translational diffusion can be best inferred from the lowest $Q=0.5$ {\AA}$^{-1}$ data (spatial scale up to $2\pi/Q\simeq13$ {\AA}). Similarly, $E_A$ associated with localized dynamics (sphere of radius $\sim$5 {\AA}) is provided by the largest investigated $Q$  of 1.3 \AA$^{-1}$. The observed $E_A$ within this small $Q$ interval increases  with  increasing $Q$, as indicated in the inset of Fig. \ref{fig.2}. There is a relatively broad distribution of $E_A$ values, whose observable limits are more less set by the lowest and highest accessible $Q$ on the neutron spectrometer. This  distribution of $E_A$, as summarized in Table \ref{tbl2}, suggests that a single Lorentzian alone cannot fully describe the entire quasi-elastic neutron (QENS) spectra. To map-out the entire observable $E_A$ values, $I(T)$ should be analyzed at each accessible $Q$ value.  Eq. \ref{eq:EIS} provides only a rough estimate of $E_A$ at the corresponding length scale probed. A more sensitive and accurate approach for determining $E_A$ for the observed diffusive process is to fit the neutron spectra with an appropriate model at each wavevector independently (in either the energy or time domains), and investigate the temperature dependence of the corresponding translational diffusion coefficient or relaxation times, as we have done below.  The parameter $A_0$ decreases with increasing $Q$, varying for example  from  0.46 at $Q=0.5$ {\AA}$^{-1}$ to 0.27 at Q$=1.3$ {\AA}$^{-1}$ in ACF-10. 

\begin{table}
\caption{ \label{tbl2} Estimate of the activation energy $E_A$  (in units of kJ/mol) of water in ACF, as obtained from fits of Eq. \ref{eq:EIS} to the integrated elastic intensity $I(T)$.  Within the $Q$-range investigated (0.5 $\le Q\le 1.3$ \AA$^{-1}$), these estimates provide lower and upper limits for $E_A$ as indicated in the inset of Fig. \ref{fig.2}. The low $Q$ yields  the characteristic $E_A$ value for long range translational diffusion (up to $2\pi/Q\simeq13$ {\AA}) and the high $Q$ gives the characteristic value for the localized dynamics (at a length scale of $\sim$5 {\AA}). The rather broad distribution of $E_A$ values indicates the limitation of the method used. To more accurately determine $E_A$, an alternative approach based on the temperature dependence of the diffusion coefficient (as done below) is needed.}
\begin{ruledtabular}
 \begin{tabular}{c c c c}
$Q$ ({\AA}$^{-1}$) & ACF-10   & ACF-20	\\
\hline
0.5	&13.6	& 16.3\\
\hline
1.3 	& 16.2  &      21.2 
\end{tabular}
\end{ruledtabular}
\end{table}

\subsection{Quasi-Elastic Neutron Scattering (QENS)}
Having determined a suitable temperature range for investigating the diffusion of water confined in the hydrated ACF samples on the spectrometer,  high statistical quality quasi-elastic neutron (QENS)  data  were collected at three temperatures: 280,  250, and 230 K.  The relevant momentum transfer range for this comparative QENS study is limited to 0.5 $\le Q\le$1.3  {\AA}$^{-1}$, $\Delta Q=0.2$ \AA$^{-1}$, to avoid  the undesirable influence of the coherent contributions of carbon at low and high $Q$'s, and of the faster rotational motions of water  that complicate the interpretation of the data.  As indicated above, Fig. \ref{fig.1} shows representative spectra as a function of  $Q$ for water in ACF-20 at 250 K. The figure also depicts the resolution function at $Q=1.3$ {\AA}$^{-1}$ taken with the same sample cooled down to 50 K (dashed black line). Each QENS spectra was normalized against the same vanadium run to correct for detector efficiency, and subsequently Fourier transformed to a self-intermediate scattering function $I(Q,t)$ using, 

\begin{equation} 
I(Q,t)\simeq\frac{1}{2\pi}\int_{-E_m}^{E_m} dE  ~\exp(iEt/\hbar)~S(Q,E)  
\label{eq.ft}
\end{equation} 
\noindent where $E_{m}$ is  $\sim$120 $\mu$eV. The instrument contributions were removed by dividing the resulting $I(Q,t)$ at each temperature by that obtained at 50 K data. The intrinsic $I_{in}(Q,t)$ were then all fitted in the time domain over a limited but reliable $t$ range, 30-750 ps. In principle, the accessible times on BASIS is a low as 5 ps and as high as 1000 ps, given the dynamics range and the energy resolution of the instrument. The narrower time domain over which the present analysis was done is one for which the translational dynamics of water are conveniently relevant on BASIS, and (2) where known systematic errors associated with complex Fourier transform of neutron data have minimal effects. Further details regarding the Fourier transform methods used here can be found in Ref. \cite{Mukhopadhyay:14}.
 
\begin{figure}
\includegraphics[width=0.8\linewidth]{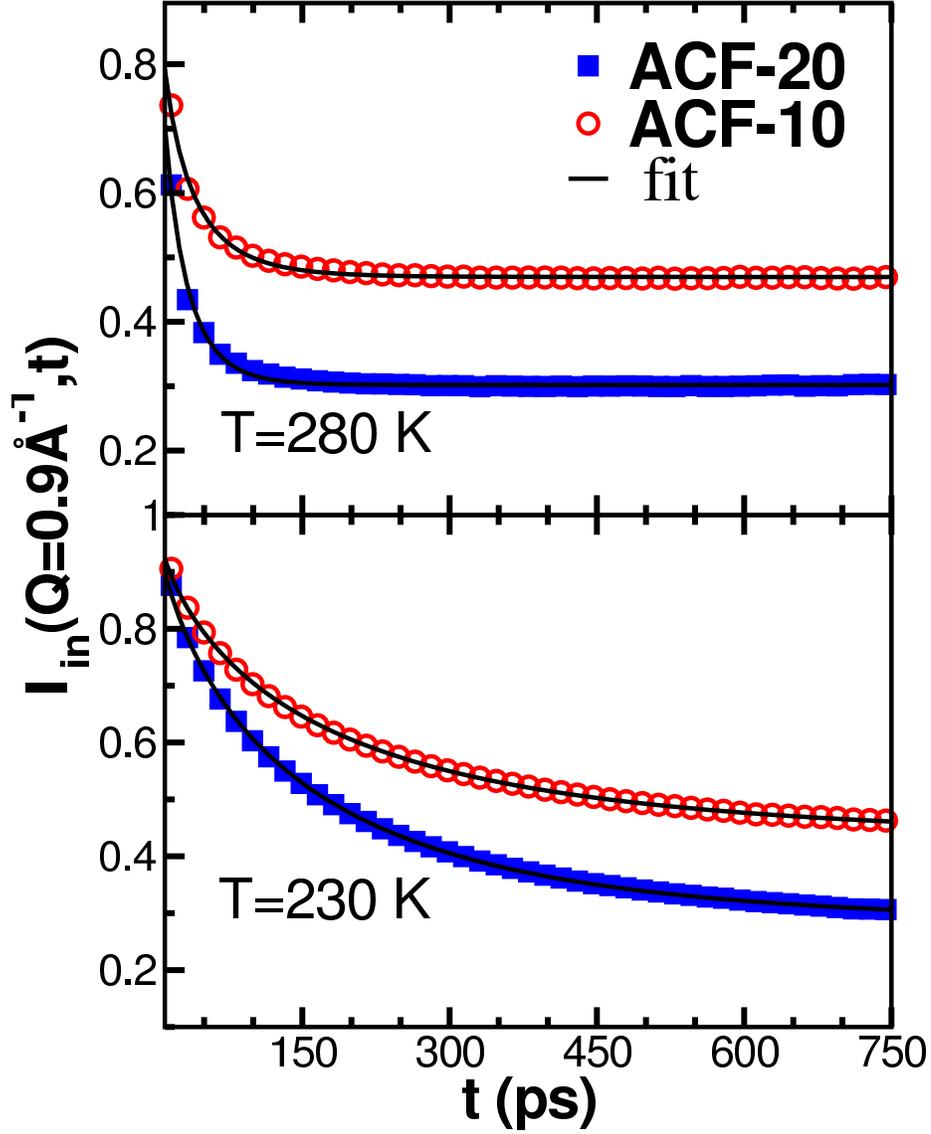}
\caption{(Color online) Intrinsic Intermediate scattering function $I_{in}(Q,t)$ of water confined in activated carbon fibers, as obtained from Fourier transforming the observed $S(Q,E)$ data, and removing resolution contributions at a selected $Q$ of $0.9$ {\AA}, as explained in the text.  The top panel compares the results derived for water in ACF-10 (open circles) and ACF-20 (closed circles) at temperature $T=$ 280 K. The bottom panel makes a  similar comparison of the data collected at  230 K. The solid black lines are model fit using the stretched exponential model, described in the text. The errorbars associated with $I_{in}(Q,t)$ are less than 1\% within the time range shown.}
\label{fig.3}
\end{figure}

\section{Data Analysis}\label{sec:DA}

Fig. \ref{fig.3} compares the resolution deconvoluted intrinsic $I_{in}(Q,t)$ for $Q=0.9$ {\AA}$^{-1}$ at two selected temperatures. These resolution independent $I_{in}(Q,t)$ could be adequately described by a Kohlrausch-Williams-Watts (KWW) stretched exponential model \cite{Williams:70} plus a time-independent elastic component,
\begin{equation}
I_{in}(Q,t)= A_0+\left(1-A_0\right)\exp \left[-\left(\frac{t}{\tau}\right)^{\beta}\right]
\label{eq.kww}
\end{equation}
\noindent where $A_0$ denotes the fraction of immobile water molecules (most notoriously known as elastic incoherent structure factor or $EISF$), $\tau$ is the relaxation time, and $\beta$ the stretching exponent. Our $A_0$ parameter is  offset from the true EISF because \lq dry' sample contribution has not been subtracted here. In our recent investigation of water in ACF-10 \cite{Diallo:15}, we found its dynamics to be more reliably captured by the KWW model, rather than the two exponential model. This KWW model is suggestive of and consistent with a heterogeneous diffusive dynamics. Typical KWW fits can be seen as black solid lines in Fig. \ref{fig.3}.  As can be appreciated, the fits clearly capture the experimental data quite reliably over a wide time range, from about 30 ps to about 750 ps. From these fits, the temperature and wavevector dependence of the three adjustable parameters $A_0$, $\tau$ and $\beta$ are determined.  By analyzing the $Q$-dependence of these parameters at each temperature,  the diffusion coefficient, and average relaxation time $\langle \tau_{\beta} \rangle$ for the confined water can be determined.  Assuming a distribution of relaxation times at  temperature $T$,  the mean value  of $\langle \tau_{\beta} \rangle$ at that temperature is thus the area $\int_0^\infty dt ~e^{(-t/\tau)^{\beta}}= (\tau/\beta) \Gamma(1/\beta)$, where $\Gamma(x)$ is the gamma function.

\section{Results and Discussion}\label{sec:res_disc}
\begin{figure}
\includegraphics[width=0.9\linewidth]{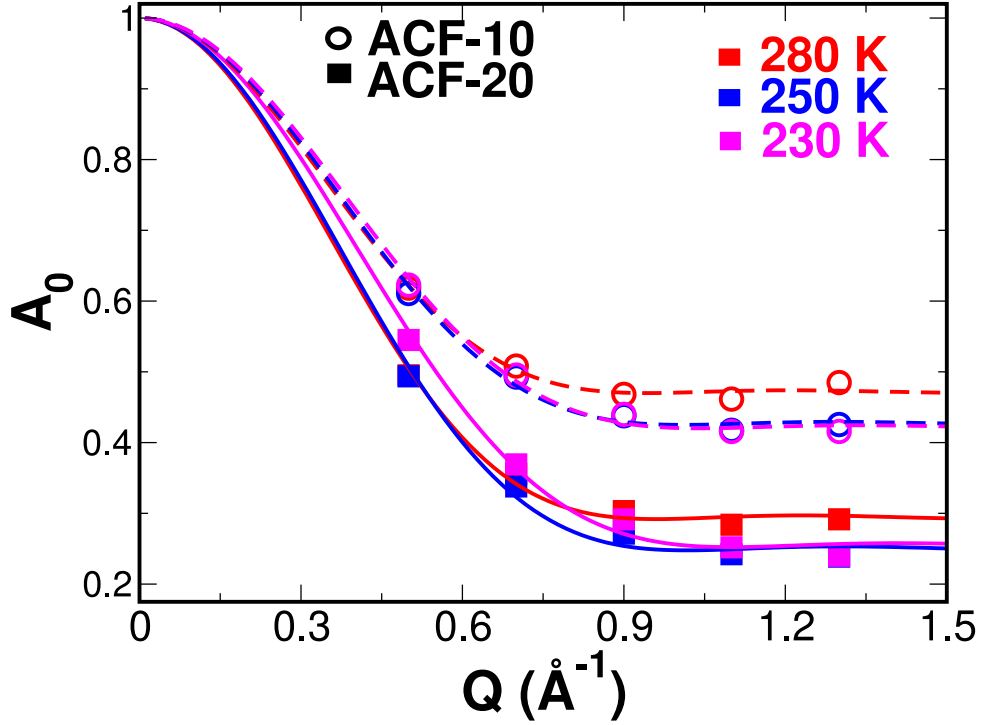}
\caption{(Color online) Temperature and momentum transfer dependence of the observed  $A_0$ (offset elastic incoherent structure factor) for ACF-10 (open circles) and ACF-20 (solid squares). }
\label{fig.4}
\end{figure}

The $Q$-dependence of the $A_0$  parameter (from Eq. \ref{eq.kww}) associated with each ACF sample is displayed in Fig. \ref{fig.4} for the three investigated temperatures. This parameter introduced above is a measure of the fraction of H atoms that do not contribute to the diffusion process within a cage of certain dimension ($\sim2\pi/Q$).  With this in mind, it is evident (based on Fig. \ref{fig.4}) that (1) the fraction of mobile water molecules is largely temperature independent (except perhaps in ACF-10), and (2) that at short distances (high $Q$), approximately 20\% more water molecules will be contributing to the observed dynamics in the large pores of ACF-20 than in the narrow ACF-10 pores, in agreement with Fig. \ref{fig.2}. To estimate the size $a_i$ (where $i=$10, 20 for ACF-10, and ACF-20 respectively) of the confining cage from $A_0$, a model fit is necessary. Using the following generic expression EISF for spherically confined motion, \cite{Bellissent-Funel:95},
\begin{equation}
A_0=f_i + (1-f_i)\left(\frac{3j_1(Q a_i)}{Q a_i}\right)^2
\label{eq.eisf}
\end{equation} 
\noindent where $j_1$ is the spherical Bessel function, the parameter $a_i$ and the actual fraction $f_i$ of immobile water molecules outside the cage are extracted at each temperature, as summarized in Table \ref{tbl3} for both samples ($i=$ ACF-10, ACF-20). These values confirm the larger fraction of immobile water molecules in ACF-10 inferred from Fig. \ref{fig.4}, and reveal an average confining cage radius for water in ACF-20 of about 4.5 {\AA} that is unaffected by temperature (less than 2\% change on cooling from 280 K to 230 K). In contrast, the corresponding cage  size in ACF-10 shrinks by as much as 16 \% for the same temperature change.

\begin{table}
\caption{\label{tbl3} Confining cage size $a_i$, and fraction $f_i$ of immobile of hydrogen atoms, obtained from fits of Eq. \ref{eq.eisf}  to the elastic incoherent structure factor  shown in Fig. \ref{fig.4}.}
\begin{ruledtabular}
\begin{tabular}{c c c c c}
$T$ (K) & a$_{10}$ ({\AA})  & a$_{20}$ ({\AA}) & f$_{10}$ (\%) & f$_{20}$ (\%)	\\
\hline
280	& 4.8(2) 	&  4.7(1) & 47 & 29 \\
\hline
250 	&   4.4(1) &  4.4(3)        & 43 & 25 \\
\hline
230 &   4.0(1) &     4.6(1)  &  42 & 25
\end{tabular}
\end{ruledtabular}
\end{table}

\begin{figure}
\includegraphics[width=0.9\linewidth]{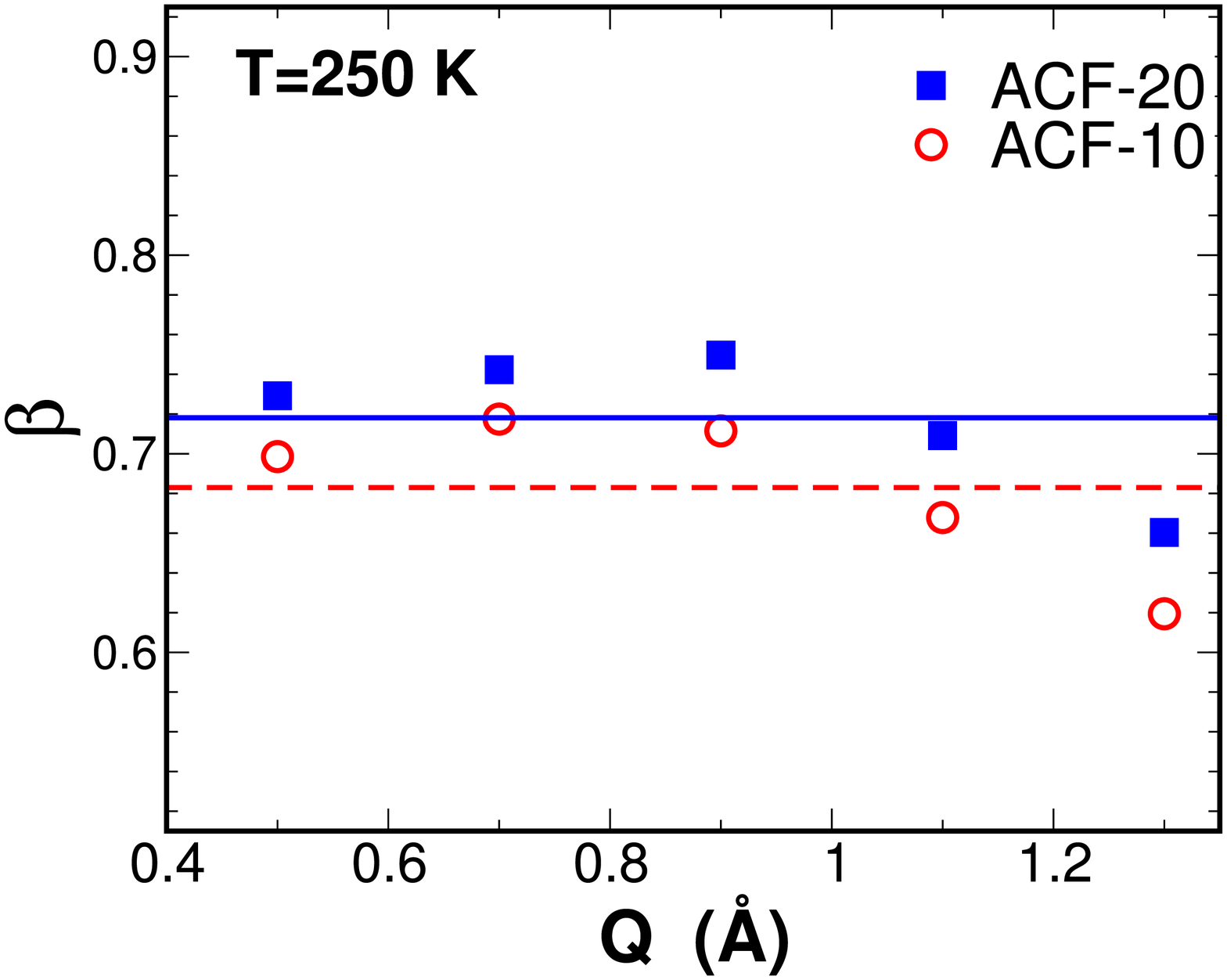}
\caption{(Color online) Best fit stretching exponent parameter $\beta$ as a function of $Q$ at $T=250 K$ for ACF-10 (open circles) and ACF-20 (solid squares), determined within 0.5\% precision. The solid and dashed lines represent the $Q$-averaged  $\langle \beta\rangle$ values for ACF-10 (0.68) and ACF-20 (0.71), respectively. In the present work, $\beta$ for each $Q$ and $T$ were used instead of $\langle \beta\rangle$, as is often the case.}
\label{fig.5}
\end{figure}

Eq. \ref{eq.kww} yields a stretching exponent $\beta$, as well as a relaxation time $\tau$ that are both temperature and wavevector dependent. While the influence of these variables on $\beta$ is rather marginal compared to that on $\tau$ and generally kept fixed to some average value in KWW fits, we here chose to let it vary with $\tau$, and $A_0$. We subsequently computed the average relaxation time $\langle \tau_\beta\rangle$. The variation of $\beta$ with $Q$ for water confined in the ACF samples at 250 K is illustrated in Fig. \ref{fig.5}. Within the temperature and $Q$-range investigated, the observed $\beta$ values fall in the range of $0.6\le\beta \le0.8$, in excellent agreement with previous findings.\cite{Mansour:02,Crupi:03,Liu:06}

\begin{figure}
\includegraphics[width=0.9\linewidth]{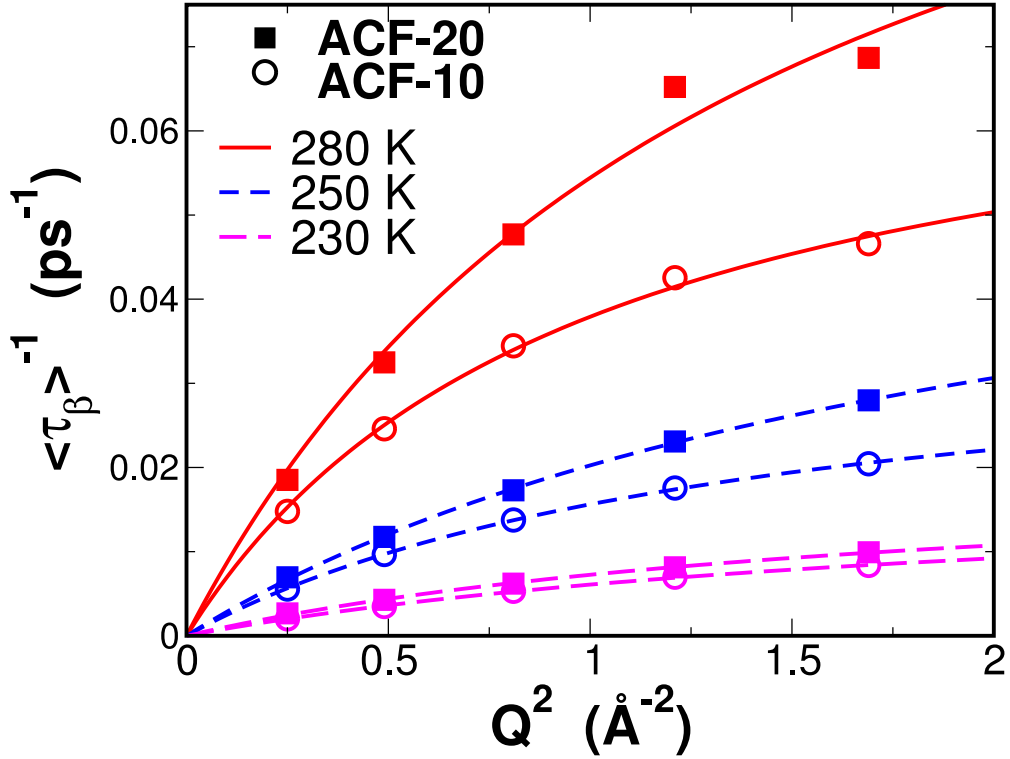}
\caption{(Color online) Inverse of the average relaxation time as a function of $Q^2$ and temperature.  The solid squares are the characteristic values obtained for ACF-20 and the open circles those for ACF-10. The Colored lines are the best representative fits using the jump diffusion model discussed in the text. The associated errorbars are between 2-9\%.}
\label{fig.6}
\end{figure}

From the $Q$-dependence of $\langle\tau_\beta\rangle$, it is possible to determine the nature of the diffusion process, whether for example it is translational (quadratic in $Q$) or rotational ($Q$ invariant) in character.  The computed values for all the three investigated temperatures  are displayed in the form of $1/\langle \tau_\beta\rangle$ versus $Q^2$ in Fig. \ref{fig.6}. The resulting $Q$-dependence suggests a translational jump diffusion, with a quadratic behavior at low $Q$'s, and a saturation to a fixed value at high enough $Q$. This model is given by,
\begin{equation}
\frac{1}{\langle\tau_\beta\rangle}=\frac{D_r Q^2}{1+D_r Q^2\langle\tau_0\rangle}
\label{eq.jump}
\end{equation} 
\noindent where $D_r$ corresponds to the average translational diffusion coefficient of water inside the pores (associated with the H-sites). The lines in Fig. \ref{fig.6} represent the best fits of Eq. \ref{eq.jump} to the experimental data, from which the temperature dependence of $\langle\tau_0\rangle$ and $D_r$ are extracted and examined below.

\begin{figure}
\includegraphics[width=0.9\linewidth]{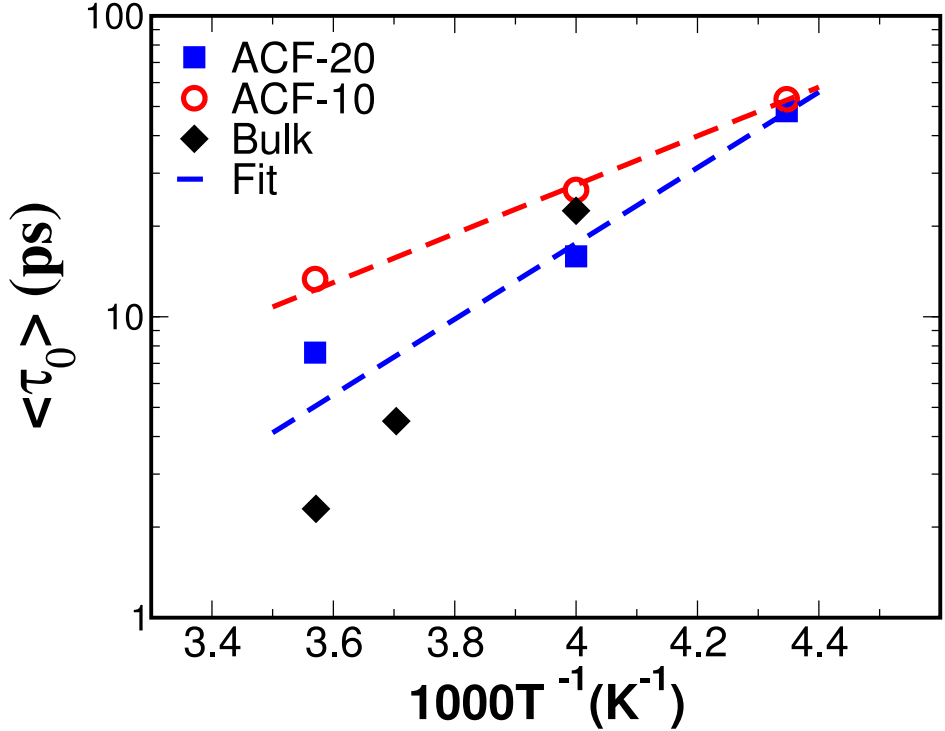}
\caption{ (Color online) Mean relaxation time $\langle\tau_0\rangle$ of water confined in ACF-10 (open circles) and ACF-20 (solid circles) plotted as a function of $1000/T$. Data for bulk water \cite{Teixeira:85,Mitra:01}, observed at comparable temperatures are also shown. Dashed lines are fits to an Arrhenius temperature dependence.}
\label{fig.7}
\end{figure}

Fig. \ref{fig.7} compares the temperature dependence of $\langle\tau_0\rangle$ for water in ACF-10 and ACF-20 with that of bulk water, plotted  on a logarithmic scale. Within the instrumental precision, our observed relaxation time of water $\langle\tau_0\rangle$ changes by a factor of $\sim$6  between  230 K and 280 K in ACF-10, compared to only 4 in ACF-20, as indicated by the different slopes of the lines in Fig. \ref{fig.7}.  Since shorter times are associated  with faster dynamics, one can immediately conclude that while the dynamics of water becomes progressively  slower as the confining pore size is reduced, the corresponding $\langle\tau_0\rangle$ values between jump-sites, are consistently larger in the smaller pores at all temperatures. 

\begin{figure}
\includegraphics[width=0.9\linewidth]{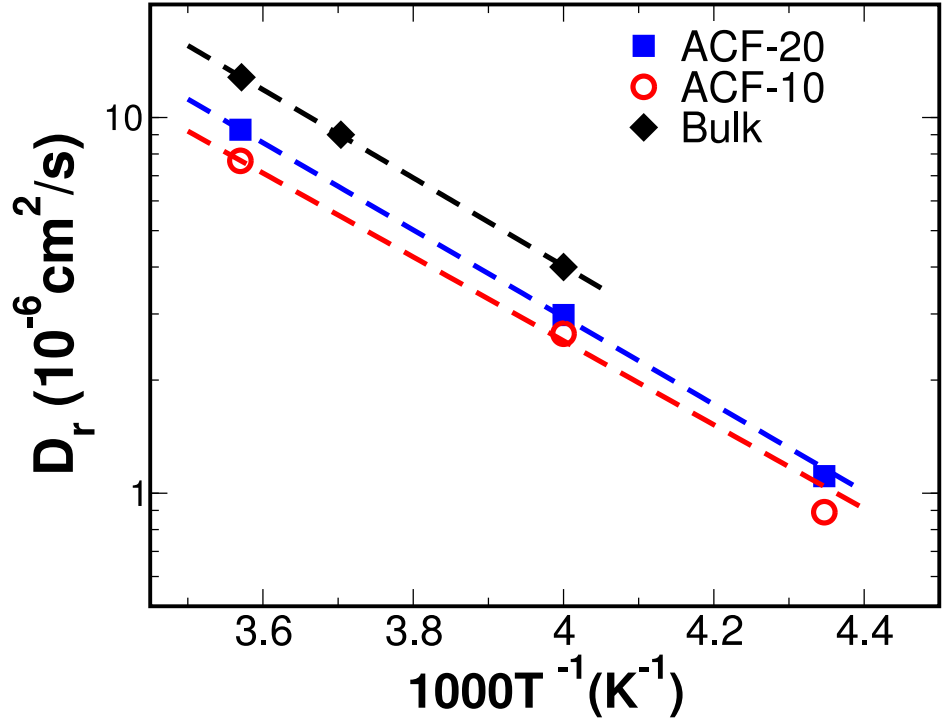}
\caption{(Color online) Net translational diffusion coefficient  as a function of $1000/T$ for ACF-10 (open circles) and ACF-20 (solid squares), as compared to that of bulk water (solid diamonds). Dashed lines are fits to an Arrhenius behavior $\sim D_0 e^{-E_a/RT}$.}
\label{fig.8}
\end{figure}

A subsequent investigation of the translational diffusion coefficient $D_r$ shown in Fig. \ref{fig.8} confirms the slowing down of the dynamics as the confining  pore dimension is reduced. The behavior of $D_r$ indicates a progressively decreasing mobility of water molecules as the confining space becomes tighter. At 280 K for example, $D_r$ is respectively 40 and 30\% lower in ACF-10 and ACF-20 than in the bulk liquid at the same temperature. As the temperature drops, $D_r$ departs significantly from its bulk value, decreasing by as much as 77\% in ACF-10 and 70\% in ACF-20 at 230 K. 

Both $D_r(T)$ and $\langle\tau_0\rangle(T)$ can be fit to an Arrhenius exponential decay $\sim A\exp(\pm E_A/RT)$, yielding the characteristic activation energy $E_A$ associated with each. The resulting fits are indicated by the dashed lines in Figs. \ref{fig.7} and \ref{fig.8}, with $E_A$ values that are summarized in Table \ref{tbl4}. The $E_A$ values inferred from $D_r(T)$ are approximately 21-22 kJ/mol, comparing well with previously reported values in porous media \cite{Liu:06,Qvist:11}. Contrary to those deduced from $\langle\tau_0\rangle(T)$, they also appear to be more uniform across the two samples, and comparable to those of bulk water  \cite{Mitra:01,Teixeira:85}. Although this  estimated bulk water $E_A$ ($\sim$22.4 kJ/mol) is significantly larger than that reported in Ref. \ref{Teixeira:85} for strictly rotational relaxations (E$_A=$7.7 kJ/mol), we find it to agree rather well with recent findings in which the corresponding molecular motions were not considered to be strictly rotational, with $E_A=$22-24 kJ/mol \cite{Qvist:11}.  Compelling arguments of why the localized water dynamics cannot be considered to be purely rotational in character can be found in Ref. \cite{Qvist:11}.

\begin{table}
\caption{\label{tbl4} Activation energy $E_{A_i}$ (kJ/mol) inferred from various parameters associated with water diffusion, where $i=$10, 20 for water in ACF-10 and ACF-20 respectively. Corresponding bulk water $E_A=E_{A_{bulk}}$ value derived from the data in Ref. \cite{Mitra:01, Teixeira:85} is shown for comparison.}
\begin{ruledtabular}
\begin{tabular}{c  c c }
Source  & $ \langle \tau_0\rangle$ & $D_r$ (10$^{-6}$cm$^2$/s)\\
\hline
$E_{A_{10}}$ (kJ/mol) &   15.5 &     21.3 \\
\hline
$E_{A_{20}}$ (kJ/mol)       &24.1  &    22.2\\ 
\hline
$E_{A_{bulk}}$    (kJ/mol)         & Ñ  & 22.4 
\end{tabular}
\end{ruledtabular}
\end{table}

Because the $E_A$ derived from the temperature dependence of $\langle\tau_0\rangle$ in Fig. \ref{fig.7} do not exhibit a clear systematic pore size dependence, we focus instead on the $E_A$ that dictates the temperature dependence of $D_r(T)$ (see Table \ref{tbl4}). The results suggest that while the overall diffusive dynamics of the water molecules slow down as the pores size or the temperature is decreased, the activation energy for H-bond breaking in water confined in porous ACF remains globally unchanged from its bulk value. 

In a recent molecular dynamics simulation (MD) work,  Chiavazzo {{\it et al.}}  \cite{Chiavazzo:14} reported a \lq universal' scaling law that can be used to interpret water transport in confined geometries. This scaling depends primarily on a single parameter, $\theta$ which is the ratio between the strictly confined water (or water influenced by the pore walls) to that of the total water in the confining pore. Obviously the larger the pore, the smaller this parameter is. They tested and validated their predictive model against some 60 cases, ranging from water in nanoporous silica to hydration layers in proteins. Based  on this proposed scaling model, the translational diffusion coefficient $D_r$ of confined water can be written  at any temperature as,
\begin{equation}
D_r(T)=\theta D_c+(1-\theta) D_{bulk}(T)
\label{eq.scaling}
\end{equation}
\noindent where $D_{bulk}(T)$ is the bulk water translational diffusion at temperature $T$, and $D_c$ and $\theta$ the temperature insensitive parameters, respectively the translational diffusion coefficient associated with the strictly confined water (\ie the water that is most strictly affected by confinement), and $\theta$ the ratio between this water and the total water in the confining media. In practice, $D_{bulk}(T)$ can be obtained from the literature, but $D_c$ and $\theta$ would be sample specific and geometry dependent.

\begin{table}
\caption{Observed temperature independent characteristic parameters of water confined in activated carbons;  the fraction $\theta$ of water molecules closest to the pore walls ($\frac{N_{bound}}{N_total}$), and its corresponding diffusion coefficient $D_c$\label{tbl5}}
\begin{ruledtabular}
\begin{tabular}{c  c c }
Parameters  & $\theta=\frac{N_{bound}}{N_{total}}$ & $D_c$ (10$^{-6}$cm$^2$/s)\\
\hline
ACF-10 &   0.41 & 0.35  \\
\hline
ACF-20 &  0.28 & 0.14   
\end{tabular}
\end{ruledtabular}
\end{table}

\begin{figure}
\includegraphics[width=0.9\linewidth]{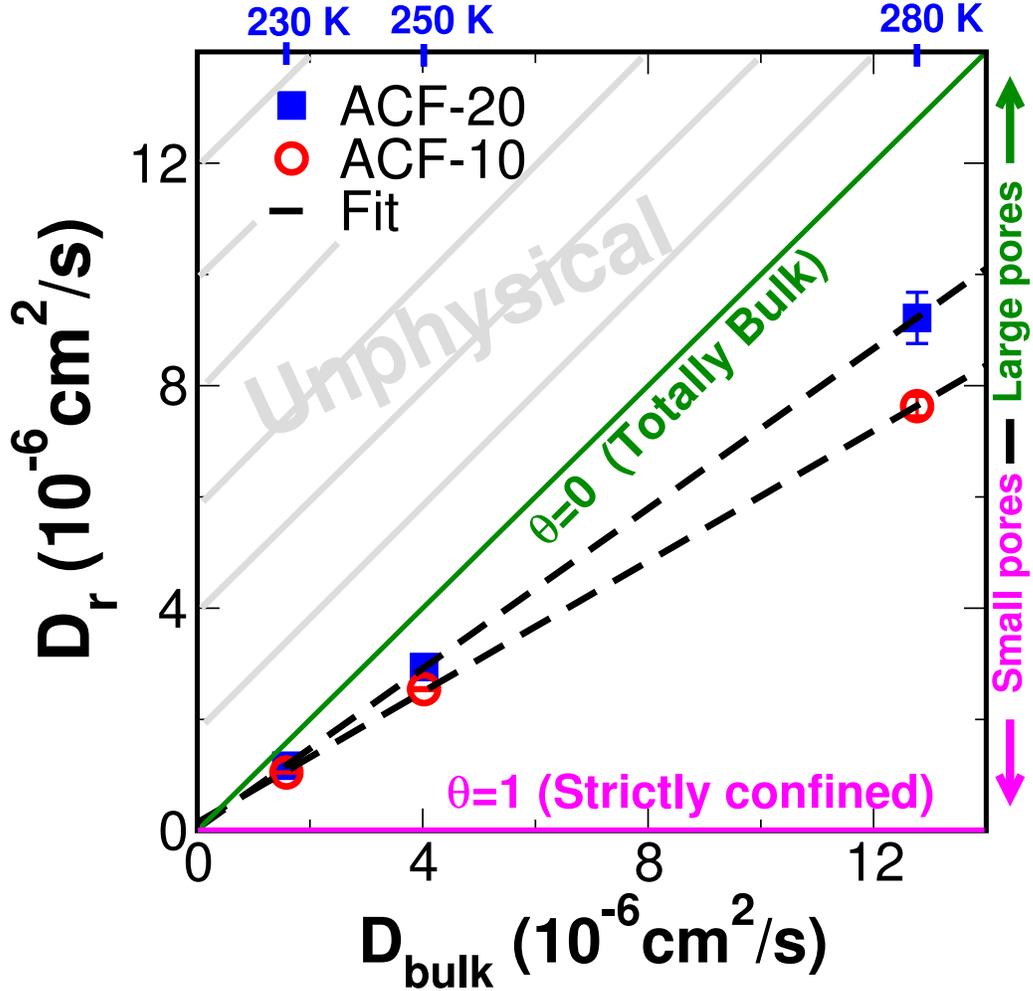}
\caption{(Color online) Translational diffusion coefficient ($D_r$) of water restricted within the pores of activated carbons at the three temperatures investigated plotted against the corresponding values for bulk water ($D_{bulk}$). The lowest $D_{bulk}$ value (230 K) is based on an extrapolation of the Arrhenius behavior of $D_{bulk}$ shown  in Fig.\ref{fig.8}. The dashed lines represents the scaling behavior $D_r(T)=\theta D_c+(1-\theta) D_{bulk}(T)$ proposed in Ref. \cite{Chiavazzo:14}, where $D_c$ is a temperature independent diffusion coefficient associated with the strictly confined water (\ie most influenced by the pore walls), and $\theta$ the ratio between this water and the total water inside the pores. The grey shaded area is not physically accessible since $D_r(T)$ cannot exceed the bulk value. The characteristic $\theta$ and $D_c$ parameters for water diffusing in ACF-10 and ACF-20 are summarized in Table \ref{tbl5}.}
\label{fig.9}
\end{figure}

Using  Eq. \ref{eq.scaling}, it is thus possible to determine the characteristic transport parameters $D_c$ and $\theta$ for water in the present ACF samples. Fig. \ref{fig.9} shows the observed $D_r(T)$ values in ACF versus known bulk water data ($D_{bulk}(T)$). The dashed lines are fits of Eq. \ref{eq.scaling} to the experimentally observed values. From the slopes and intercepts of these lines, $\theta$ and $D_c$ can be extracted. The observed values for each sample are listed in Table \ref{tbl5}. Fig. \ref{fig.9} also shows the limiting situations by the strictly confined water ($\theta=1$) and pure bulk water ($\theta=0$). The grey shaded area is inaccessible and corresponds to the unphysical  condition $D_r(T)\ge D_{bulk}(T)$. 

In our opinion, the temperature region over which Eq. \ref{eq.scaling} is applicable may depend on both the sample and the characteristics of the neutron instrument used (resolution and dynamics range). Based on the current data, we estimate this range to be between $T_{min}\sim$ 220 K and $T_{max} \sim$290 K on the BASIS spectrometer. The maximum value is primarily set by the accessible time window on the instrument, and the lower limit varies based on a combination of instrumental resolution, and the physics of water itself (confining pore size, possible phase transition related to thermodynamics anomalies \cite{Liu:04} \etc). It may however be more prudent to set  $T_{min}$ above at least 250 K, the lowest temperature at which bulk supercooled water can be easily achieved and for which neutron scattering work has been fairly well documented \cite{Qvist:11,Mitra:01,Teixeira:85}. The law requires thus that  the temperature range considered be one in which $D_c$ and $\theta$ do not vary with temperature. This suggests  a temperature interval in which the effective potential well in which water sits in the confining pore remains greater than the thermal kinetic energy.

In the current study, the translational diffusion coefficient for bulk water at 230 K shown in Fig. \ref{fig.9} (lowest value where $D_r$ for ACF-10 and ACF-20 appears to overlap) was estimated by extrapolating the temperature dependence of $D_{bulk}(T)$ above 250 K down to  230 K. This approximation is off course by no means exact since $D_r$ of water is known to be rather super-Arrhenius \cite{Teixeira:85}, but it is one that yields a reasonable estimate for the present relatively small temperature interval. Excluding the data at 230 K does not affect our findings. In fact, the remarkable observed linearity in Fig. \ref{fig.9} suggests that a single temperature point would be sufficient to predict the overall temperature behavior, and the characteristic parameters $D_c$ and $\theta$. This observation warrants further investigation. 

\section{Conclusions}\label{sec:con}

In the present study, the molecular dynamics of water adsorbed in microporous activated carbon fibers have been investigated for two different pore sizes, and compared with bulk water dynamics. The quasi-elastic neutron scattering data of ACF-10 ($\sim$12 {\AA}) and ACF-20 ($\sim$ 18 {\AA}) reveal a retardation of the water dynamics when either the pore dimension or the temperature is reduced.  The observed translational diffusion coefficients can be adequately described by a recently proposed scaling law for water diffusion in nano-confined geometries, yielding intrinsic characteristic parameters capable of predicting water mobility in each of these two specific porous carbon materials at any temperature over a wide temperature range where the law remains valid. The significance of this scaling law to neutron experimenters, who far too often  struggle to complete a series of temperature scans within a limited beamtime allocated by the neutron facilities, cannot be underestimated. The present findings suggest that for a given nanoporous media, knowledge of the water dynamics  at 2 temperatures (3 to confirm accuracy, as we have done here) is all that is required for predicting its diffusion at other temperatures. To further test the universality of this law,  additional work is however needed.  Future research could involve other well studied porous media for which the dynamical properties of water are well known (MCM-41 for example). It will also be interesting to further investigate the temperature range over which  the method can be reliably applied using other neutron spectrometers with different energy resolution and dynamics range. 

\section{Acknowledgments}
The author is thankful to R. Goyette,  R. Mills,  R. Moody, and M. Rucker at the Spallation Neutron Source (SNS), Oak Ridge National Laboratory (ORNL) for their excellent technical support  during the measurements.  He is also indebted to Eugene Mamontov for many stimulating scientific discussions and a wonderful collegiality over the last few years.  He thanks both E. Mamontov and A. Kolesnikov for their critical reading of the manuscript. The author owes much gratitude to M. Sliwinska-Bartkowiak, A. A. Chialvo, L. Vleck  and J. S. Hayes for recent collaborations on the diffusion properties of water confined inside related nanoporous carbon materials, which have motivated the present work. The use of the Mantid software package \cite{Arnold:14} is gratefully acknowledged. This work at ORNL's SNS is sponsored by the Scientific User Facilities Division, Office of Basic Energy Sciences, US Department of Energy.  

\clearpage 
{\footnotesize
$^{*}$ Kynol$^{TM}$ is a registered trademark of Gun Ei Chemical Industry Co., Ltd. for novoloid fibers and textiles.

\bibliographystyle{unsrt}
\bibliography{../bibs/abbrevs,../bibs/Gen,../bibs/A_F,../bibs/G_P,../bibs/R_Z}
}
\end{document}